\documentclass[sigconf]{acmart}

\AtBeginDocument{%
  \providecommand\BibTeX{{%
    \normalfont B\kern-0.5em{\scshape i\kern-0.25em b}\kern-0.8em\TeX}}}

\acmConference[IMX '20]{Woodstock '18: ACM Symposium on Neural
  Gaze Detection}{JUNE 17 -19 , 2020}{Barcelona, Spain}
\acmBooktitle{Woodstock '18: ACM Symposium on Neural Gaze Detection,
  June 03--05, 2018, Woodstock, NY}
\usepackage{amsmath}
\usepackage{colortbl}
\usepackage{tikz}
\usepackage{booktabs}
\usepackage{textcomp}
\usepackage{soul}
\usepackage{amsmath}
\usepackage{balance}   
\newcommand{\sys}{Sifter}

 \setcopyright{none} 

\begin{document}


\title[\sys{}]{\sys{}: A Hybrid Workflow \\ for Theme-based Video Curation at Scale}

\author{Yan Chen}

\affiliation{%
  \institution{University of Michigan}
  \city{Ann Arbor}
  \state{Michigan}
}
\email{yanchenm@umich.edu}

\author{Andr\'{e}s Monroy-Hern\'{a}ndez}
\affiliation{%
 \institution{Snap Inc.}
 \streetaddress{2025 1st Ave}
 \city{Seattle}
 \state{WA}
  \country{USA}
  }
\email{amh@snap.com}
\author{Ian Wehrman}
\affiliation{%
 \institution{Snap Inc.}
 \city{Santa Monica}
 \state{CA}
 \country{USA}
}
\email{iwehrman@snap.com}

\author{Steve Oney}
\affiliation{%
  \institution{University of Michigan}
  \city{Ann Arbor}
  \state{MI}
  \country{USA}
}
\email{soney@umich.edu}

\author{Walter S. Lasecki}
\affiliation{%
  \institution{University of Michigan}
  \city{Ann Arbor}
  \state{MI}
  \country{USA}
}
\email{wlasecki@umich.edu}

\author{Rajan Vaish}
\affiliation{%
 \institution{Snap Inc.}
 \city{Santa Monica}
 \state{CA}
 \country{USA}
 }
\email{rvaish@snap.com}


\renewcommand{\shortauthors}{Chen, et al.}


\begin{abstract}
User-generated content platforms curate their vast repositories into thematic compilations that facilitate the discovery of high-quality material.
Platforms that seek tight editorial control employ people to do this curation, but this process involves time-consuming routine tasks, such as sifting through thousands of videos. 
We introduce \sys{}, a system that improves the curation process by combining automated techniques with 
a human-powered pipeline that browses, selects, and reaches an agreement on what videos to include in a compilation. 
We evaluated \sys{} by creating 12 compilations from over 34,000 user-generated videos. \sys{} was more than three times
faster than dedicated curators, and its output was of comparable quality. 
We reflect on the challenges and opportunities introduced by \sys{} to inform the design of content curation systems that need subjective human judgments of videos at scale.

\end{abstract}

\maketitle





\section{Introduction}

\begin{figure}[t!]
 \begin{center}
    \includegraphics[width=0.5\textwidth]{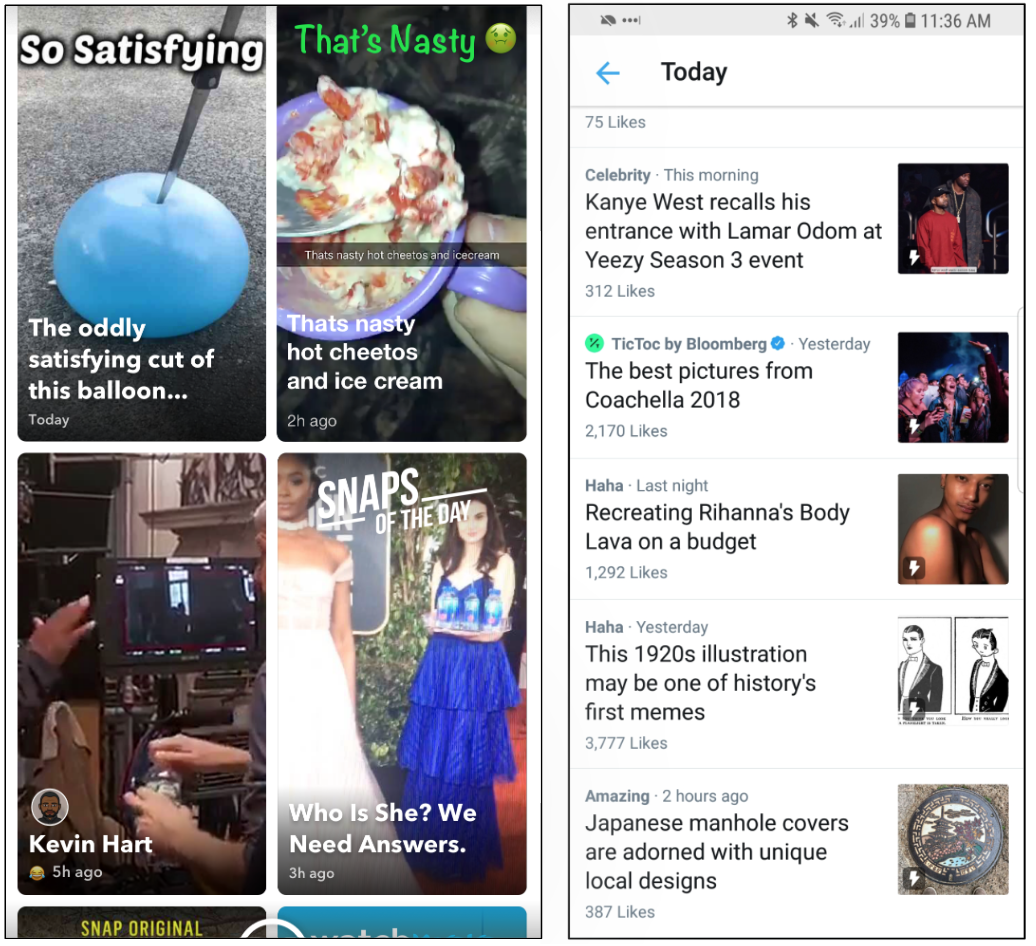}
  \end{center}
  \caption{Screenshot from 2018 of some curated content from users' posts in Snapchat, e.g., ``So Satisfying: The oddly statisfying cut of this balloon...'' (left),
  and Twitter, e.g., ``The best pictures from Coachella 2018'' (right).}
  \label{fig:story_name}
\end{figure}

Every day, millions of people around the world create, share, and consume short videos on platforms like Snapchat, TikTok, and Douyin.
These platforms use a variety of curation approaches to help their users discover high-quality and recent (``fresh'') content.
These approaches leverage artificial intelligence (AI), user-sourcing, or dedicated curators~\cite{Curry2010}. AI techniques rely on algorithmic aggregation and the ranking of relevant content based on metadata, such as tags~\cite{ortis}. 
These approaches are scalable but limited in their capacity to identify content attributes that require subjective assessments and nuanced cultural understanding. User-sourcing approaches rely on end-users' votes or ``likes'' to identify high-quality popular content, such as on Reddit ~\cite{reddit}. These approaches are also scalable, but have the potential to silence minority opinions or to be dominated by content manipulation strategies like ``brigading''~\cite{chipidza2016negative}. 
Lastly, curator-based approaches rely on staff curators to identify and organize compelling content, such as on Snapchat's Discover~\cite{discoverstory} or Twitter's Moments~\cite{twittermoment} (Fig.~\ref{fig:story_name}). 
These approaches give platforms editorial control and overcome machines' inability to make subjective assessments and prevent adversarial users from manipulating content selection, but are limited by scale~\cite{Curry2010}. 
Specifically, it is difficult to scale curators' ability to find appropriate content from a corpus of videos that is large and rapidly growing---on Youtube, for example, over 500 hours of video content is uploaded every minute.

\begin{figure}[t]
  \includegraphics[width=0.5\textwidth]{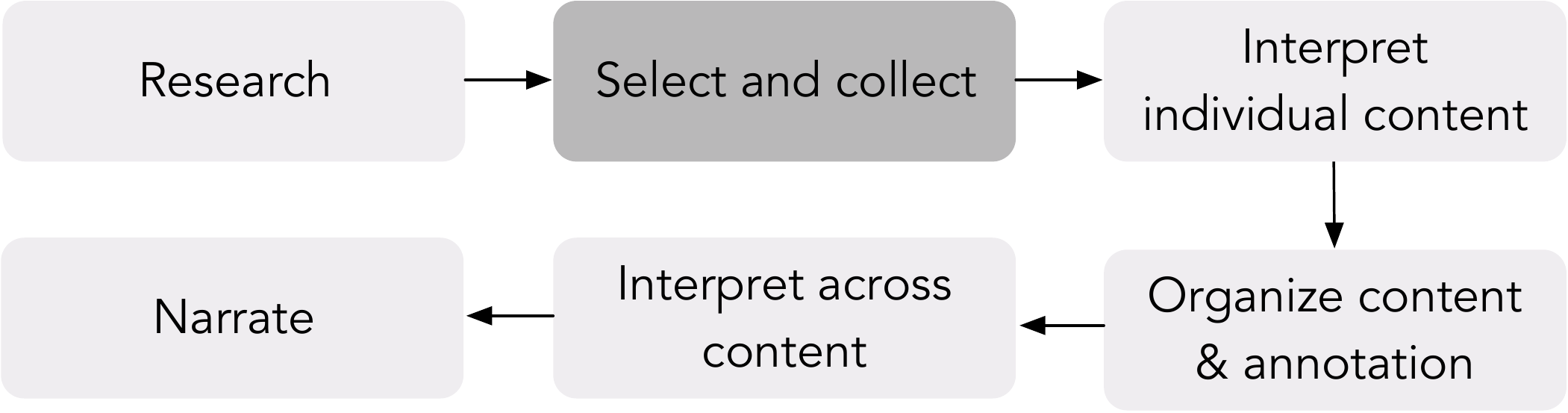}
  \caption{Common curation pipeline. We focused on the ``Select and collect'' step (gray).}
  \label{curation_pipeline}
\end{figure}

In this paper, we introduce \emph{\sys{}} to scale the third type of curation strategy (dedicated curators). 
\sys{} combines automated video processing techniques and crowdsourced human expertise to provide on-demand assistance to dedicated video curators in the process of selecting and collecting content (i.e., the ``select and collect'' phase in Fig.~\ref{curation_pipeline}).
In this phase, curators have to rapidly browse through large ``fresh-content'' corpora to collect just enough raw material that might fit a coherent narrative~\cite{wolff2013curation,askalidis2013theoretical}, or theme (e.g., ``magic tricks'', or the movie Lion King). 
As the corpora often have more appropriate (e.g., interesting, relevant) materials than needed, curators do not have to exhaust all the items.

This setting makes our problem unique, but daunting for three main reasons.
First, despite being short, videos often take more time to consume and interpret than other media, like images, as they contain multimodal signals (e.g., visual, audio, text caption). 
As a result, curators may have to watch the videos multiple times to grasp the essence, which extends the task time.
Second, the corpora often contain many unqualified videos that are distracting, further slowing curators down.
Although automated video analysis techniques have become promising, machines are still limited on assessing engaging or novel content recognition due to the social nature of content~\cite{ackerman2000intellectual}, and the difficulties in obtaining labelled training data~\cite{fu2015robust, yeung2018every,fu2018recent}. 
Third, existing crowd workflows might help more accurately assess video content than machines can by leveraging human capacity, but they often require workers to reach a specified agreement level either in parallel, or through sequential refinement~\cite{bernstein2010soylent,cranshaw2017calendar,merritt2017kurator,dai2011artificial,mohanty2019photo,zhang2018evaluation}.
Agreement can be difficult when the goal is to quickly extract a small set of videos from large video corpora where many are qualified, because, while they may not disagree with each others' selections, the sheer volume of content may result in workers selecting non-overlapping sets of responses.


\sys{} addresses the above three challenges in the following ways. 
For the first challenge, we designed a custom interface for \sys{} (Fig.~\ref{final_ui}) to make browsing videos more efficient.
For the second challenge, \sys{} automatically refines and reduces the dataset by filtering out the obviously unqualified videos (e.g., too dark, noisy) using video processing techniques.
For the third challenge, we aimed to increase the overlapping sets among workers' selections while smoothing out individual workers' biases.
So we developed a human-powered pipeline that further refines and reduces the output, and then draws a dozen or two qualified videos by agreement.
Together, \sys{} consists of a three-stage pipeline:
\begin{enumerate}
  \item \sys{} leverages automation and video processing techniques to filter out low-quality videos from a large set.
  \item \sys{} leverages human workers to rapidly select and collect thematically relevant and interesting videos.
  \item \sys{} leverages separate groups of workers to make selections from the refined set of videos, and reach an agreement.
\end{enumerate}

We evaluated \sys{} using publicly-available content from Snapchat's ``Our Stories,'' which are ``collections of Snaps submitted from different Snapchatters throughout the community'' that are ``collected and categorized to capture a place, event, or topic from different points-of-view'' \cite{ourstory}. 
More specifically, we used the themes and keywords that were used to curate 12 published compilations by staff editors or by Team Snapchat~\cite{snapsupport}, and we used \sys{} to select and collect videos to attempt to recreate those compilations.
Furthermore, we recruited three external dedicated content evaluators to assess the compilations, and we found that it took \sys{} less time ($\mu = .71$min, $\sigma = .41$min) to pick a video than the staff curators ($\mu = 2.35$min,  $\sigma = 1.49$min, $p < .0014$), and with no discernible differences in quality for eleven of the twelve compilations. 
Our findings aim to inform the design of systems that rely on subjective human judgments at scale. 


In this research, we make the following contributions:

\begin{itemize}
    \item The design of a human computation workflow that rapidly identifies high-quality and thematically-relevant videos from large sets of user-generated videos,
    
    \item \sys{} A system that instantiates our approach and helps curators focus on creative tasks by handing off routine ones,
    \item An evaluation with over 34,000 videos showing that \sys{} can help find videos of comparable quality to professional curators, but more quickly,
    \item Ethical guidelines for the adoption of Sifter.

\end{itemize}

\section{Related Work}

Two key challenges with curation at scale are the large amount of video content, and the subjective aspects of content assessment. In this section we discuss prior relevant work and potential gaps that \sys{} can fill.

\subsection{Automation Techniques}
One intuitive first approach for scaling up video curation is to use automated video processing techniques. Prior work in video analysis has explored methods to automatically detect activity~\cite{bandla2013active}, measure ``interestingness''~\cite{jiang2013understanding}, identify complementary content~\cite{bailer2017learning}, and even assess the level of creativity in a video~\cite{redi20146}. However, these attempts to subjectively understand video content are still at an early stage and prone to algorithmic bias~\cite{barocas2017fairness,barocas2017problem}. 


User-generated content platforms often use simpler, but more reliable techniques to assist the video curation process, such as grouping videos by using user activity logs (e.g. clicks) and metadata (e.g., title entered by user who uploaded a video).
These approaches have several limitations. First, using logs of user activity relies on exposing behavioral analysts to video corpora in order to collect behavior data. This might not be feasible if a platform strives for tighter editorial control, and wants to shield its users from unvetted content. Second, user-generated metadata itself is not always accurate or detailed enough to understand the content of a video (e.g., videos with a caption like ``Best Day'').

As we describe in the next section, we chose to use some of these metadata automation techniques, but did not rely on them alone. This gave us scalability benefits without requiring us to compromise on quality.

\subsection{Human-powered Video Analysis}
Current video curation practices rely mainly on humans to set the criteria used for selecting high-quality content.
These criteria are largely dependent on the available data and curators' tastes. 
By browsing the videos returned from a search query, curators constantly discover new contextual information and reshape the desired final video compilations in their mind.
This complex selection model, which is confined to the curator's mind, can be difficult for even the curator to precisely articulate.

Prior work has used crowdsourcing techniques for visual analysis, but has mostly focused on object or event recognition tasks. 
For instance, systems like Glance and Legion:AR~\cite{lasecki2014glance,lasecki2013real} leverage the crowd to identify events in a set of long videos in real-time.
Similarly, Zensors~\cite{laput2015zensors} and CrowdAR~\cite{salisbury2015crowdar} use the crowd to help alert end-users when certain events or objects occur in a live streaming video.
Shamma el al. proposed a community-supervised technique that leverages online users and machine learning for image selections~\cite{shamma2016finding}.
We build on this previous work and shift our focus to assessing the subjective attributes of videos, such as identifying whether or not a video is interesting.

Crowds are not only called upon for object recognition tasks: prior work also explored how to enable crowds to identify interesting content in a large corpus of video data. For example, Kim et al. analyzed videos of students interacting with MOOCs, to find which content sparked confusion or engagement~\cite{kim2014understanding}.

Similarly, Carlier et al. worked on identifying regions of interest within a video by analyzing log data that showed users' zooming interactions~\cite{carlier2010crowdsourced}. 
These studies, however, rely on user interaction data, which may not always be available.


\begin{figure}[t]
    \centering
    \includegraphics[width=0.5\textwidth]{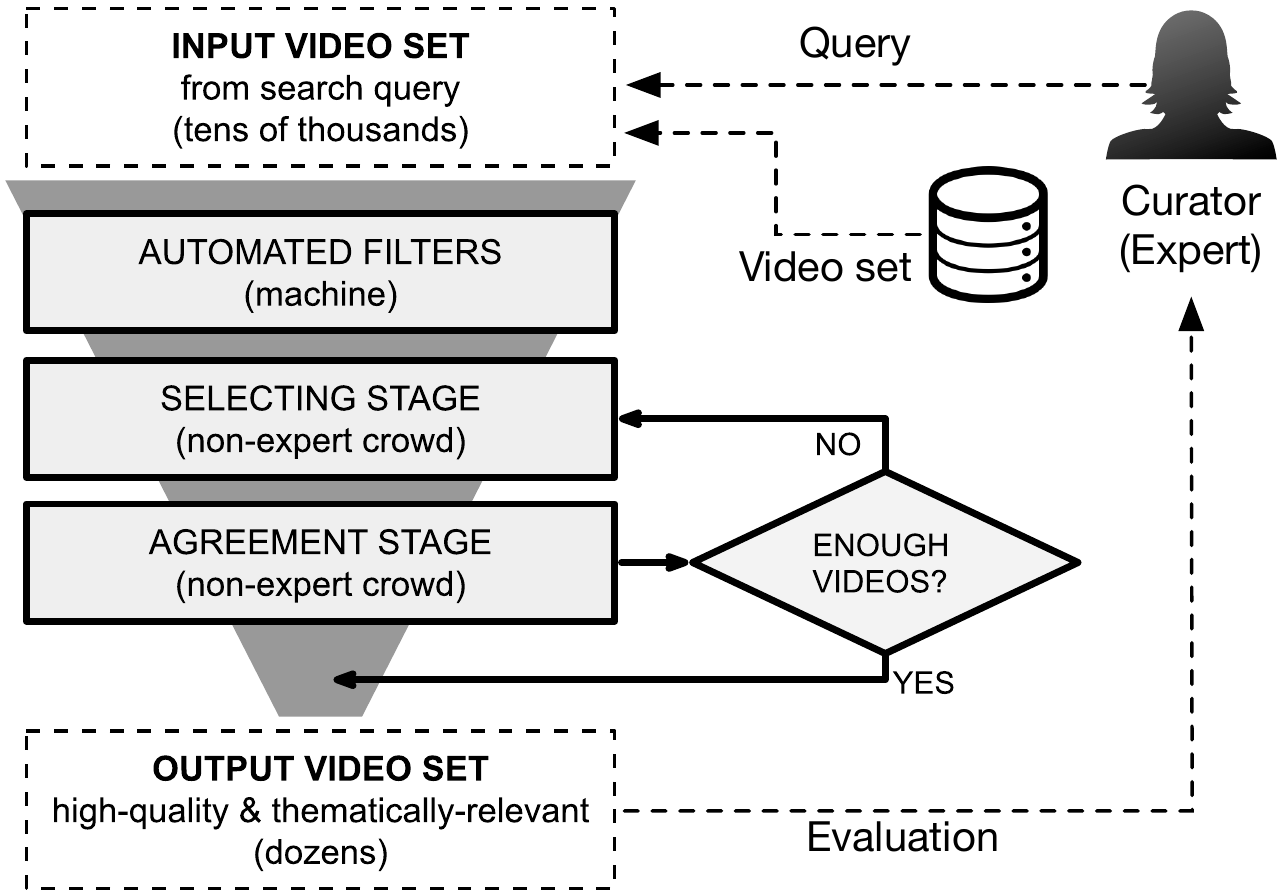}
  \caption{\sys{} pipeline.}
  \label{sys}
\end{figure}

\section{Sifter}
We created \sys{} to address the challenges of scale and subjectivity by combining human and machine computation. 
\sys{} uses video processing and human computation techniques to help scale the video curation process for a given theme (second stage in Fig.~\ref{curation_pipeline}).

The system enables staff curators to delegate the time-consuming and monotonous tasks of sifting through thousands of videos that vary in quality and selecting a small set of high-quality and thematically relevant videos. 
By high-quality (HQ), we mean videos that might capture viewers' attention and engagement, and by thematically-relevant (TR), we mean those that are well-suited for a collection of a particular topic. 

\sys{} addresses these two challenges in the following ways:

\begin{enumerate}

\item \textbf{Scale.} \sys{} addresses the challenge of scale by first leveraging automated video processing techniques (Table~\ref{table:automated_filtering}) to identify HQ videos. 
The guidelines for identifying HQ videos are derived from prior experience with videos on the platform and the existing literature (inline citations in Table~\ref{table:automated_filtering}).
Then we propose a human-powered pipeline added to the automated filter.
This workflow was derived from analyzing human workers' performance in our pilot studies.

\item \textbf{Subjectivity.} \sys{} addresses the need for subjective interpretation by relying on human workers to execute instructions.
We evaluate this method by comparing human workers' results with those of staff curators (e.g., using curators' prior search keywords).
\end{enumerate}


\sys{} includes two parts: the sifting pipeline, and the user interface. We first explain \sys{}'s three-stage pipeline, and then describe our iterative user interface design process.

\subsection{Sifting Pipeline}

The pipeline of \sys{} (Fig.~\ref{sys}) consists of three stages: 
\begin{enumerate}
\item R1: Automated filters aimed at removing trivial low-quality data for the purpose of time saving.
\item R2: Selection stage of human-worker filters quickly refine a subset of videos.
\item R3: Agreement stage of human-worker filters get multiple people's perspectives on the final decisions.
\end{enumerate}

On a high level, the pipeline works as follows. 
When the search results are returned from the public Snap post database, which often contains thousands of videos, the pipeline will first automatically process these videos to filter out the low-quality videos. 
Then the output of R1 is a quality-refined set of videos which will then be sent to a human-powered pipeline (R2,R3) for further review in a randomized order.
The human-powered pipeline can happen almost simultaneously for each stage in the long term.
The method is used as follows: the videos selected from workers in R2 are pushed to R3 in real time. 
We assign the next group of workers to R3 when the number of videos in the R3 pool reaches a threshold determined by how long we want to keep the workers in the R3 pool. This design helps us to streamline the final set of videos such that the staff curators can see more quickly what videos are selected.  


Instead of using fixed values, \sys{}'s pipeline parameterizes the input and output of each step, as well as the number of needed workers.
In both R2 and R3, workers use the aforementioned user interface and the number of videos and workers in the pool can vary.
In our final evaluation, the values we used for the parameters were derived from our pilot studies, which were conditioned on the case that we want \sys{} to generate 10-20 refined videos for each compilation (customized for the platform).
This design process and the structure of the \sys{} pipeline are generalizable as we evaluate it with a large number of videos. 
In this section, we discuss the pipeline design and implementation. 

\begin{table*}[t]
\centering
\resizebox{\textwidth}{!}{%
\begin{tabular}{p{3cm}|p{8cm}|p{8cm}}
\hline
\textbf{Filter} &  \textbf{Sifter Implementation} &  \textbf{Motivation} \\ \hline
Shorter than 3s & Use video duration property & Avg. minimum duration from published compilations   \\ \hline
Small pixel differen-\\ces between frames  & Use systematic sampling to extract five frames and compute their pixel differences in the center 200px X 200px (most important part), if more than three frames have a difference of less than 1,000 we remove the video). & Prior work showed that videos with motion of objects  (e.g., human) can be more engaging than static display (e.g., text)~\cite{guo2014video}.
\\ \hline
Low aesthetics score & Compute the average colorfulness scores~\cite{hasler2003measuring} for the same five frames and remove those videos with a score that is below a threshold we derived from a training data set of published compilations. & 
Prior work showed that video aesthetics impact engagement~\cite{chapman1997models} and that color in particular is highly correlated with aesthetics.
\\ \hline
From same session & Use video metadata to keep only one posted video from a user (we picked the first in our experiments) and eliminate the rest that are posted within the next 120s. More advanced techniques can be applied to select the best one. & Using multiple videos from the same scene and the same person would reduce the diversity of the compilation which has been shown to lower the engagement.
\\ \hline
\end{tabular}%
}
\caption{A list of automated filters, how we implemented them in \sys{}, and the motivations of using them.}
\label{table:automated_filtering}
\end{table*}

\subsubsection{R1: Automated Filters}

R1 includes four automated filters (Table~\ref{table:automated_filtering}) which take a set of a few thousand videos retrieved from a public Snap posts database as input, and outputs a refined set of videos by filtering out hundreds of unqualified videos. 
We implemented these filters by using video properties (e.g. duration) and the OpenCV library (\url{https://opencv.org/}). Table~\ref{table:automated_filtering} presents the implementation details.
 
\subsubsection{R2: Selection stage}
In this stage, a group of workers review the output videos from R1 and identify a small set of them that are high-quality and thematically relevant. 
The number of needed workers for each compilation is a parameter that depends on the number of output videos from R1. 
Based on a set of pilot studies, we found that workers perform optimally when reviewing up to around 1,000 videos and are asked to select up to about 100. 
We derived these threshold values based on observation of quality and workers' efficiency (duration of selection process).
Other researchers could derive suitable values for these parameters by exploring the trade-offs among quality, cost, and time based on their own needs. 

\subsubsection{R3: Agreement stage}
In R3, a different group of workers is given the same UI but with the output videos from R2.
The purpose of this stage is to smooth out the differences of how people interpreted the instructions and performed the task in R1.

Two workers were assigned per compilation in this final evaluation stage. The resulting output is a set of videos that both workers in R3 selected. In other words, the videos that have unanimous consent among all three workers from R2 and R3 are included in the output.

We designed R3 with two considerations.
First, the quality of the input videos (from R2) is higher than those in R2 (from R1).
With the assumption that a video with higher quality would require longer attention to review, we decided to have workers select fewer videos than in R2. 
Second, our pilot study results suggested the agreement rate is often about 40--50\% between two workers when the number of selected videos is 30 and the total is 100. 
This relatively low level of agreement is due to the fact that there are more qualified videos than needed. 
Thus, we decided to match those numbers given that the final number of videos selected is 10.
Again, other researchers may find different values for these parameters if the number of needed videos is different. 

\subsection{User Interface}

The user interface (UI) we designed for the human-powered stages of \sys{} is intended to give concrete instructions that are readily interpretable and enable the completion of open-ended tasks, i.e., find HQ and TR videos.
Designing a user interface that facilitates efficient video selection is a non-trivial process.
The large number of permutations of UI parameters  (e.g., video playback speed, number of videos per page, autoplay on/off, sound on/off,  etc.) makes it challenging to find and test a single UI for efficient video sifting. 
Pavel et al. developed a video review tool aimed to help with the video review process~ \cite{pavel2016vidcrit}. 
However, this tool focused on frame-by-frame editing of a single long video, rather than a large number of small videos. 
We addressed these concerns by first conducting a series of small studies to compare the outcome (in terms of speed) of different combinations of the major UI elements and of variants of their parameters.
Inspired by prior work \cite{krishna2016embracing}, we implemented a unique time-enforced interface that provided video auto-looping (max 10 seconds) of all eight videos on the task page at once, enabling workers to make rapid decisions. 

We iteratively designed \sys{} with two considerations in mind: the UI should be easy to use with a minimal learning curve (for speed), and workers should have enough context during the task to make a clear judgment (for quality). These features emerged from observing workers using the system during formative studies and a series of user studies where we tested each UI component one at a time.  The final interface consisted of a landing page with instructions and a task page with five main components: 

\begin{figure}[h]
    \centering
  \includegraphics[width=0.5\textwidth]{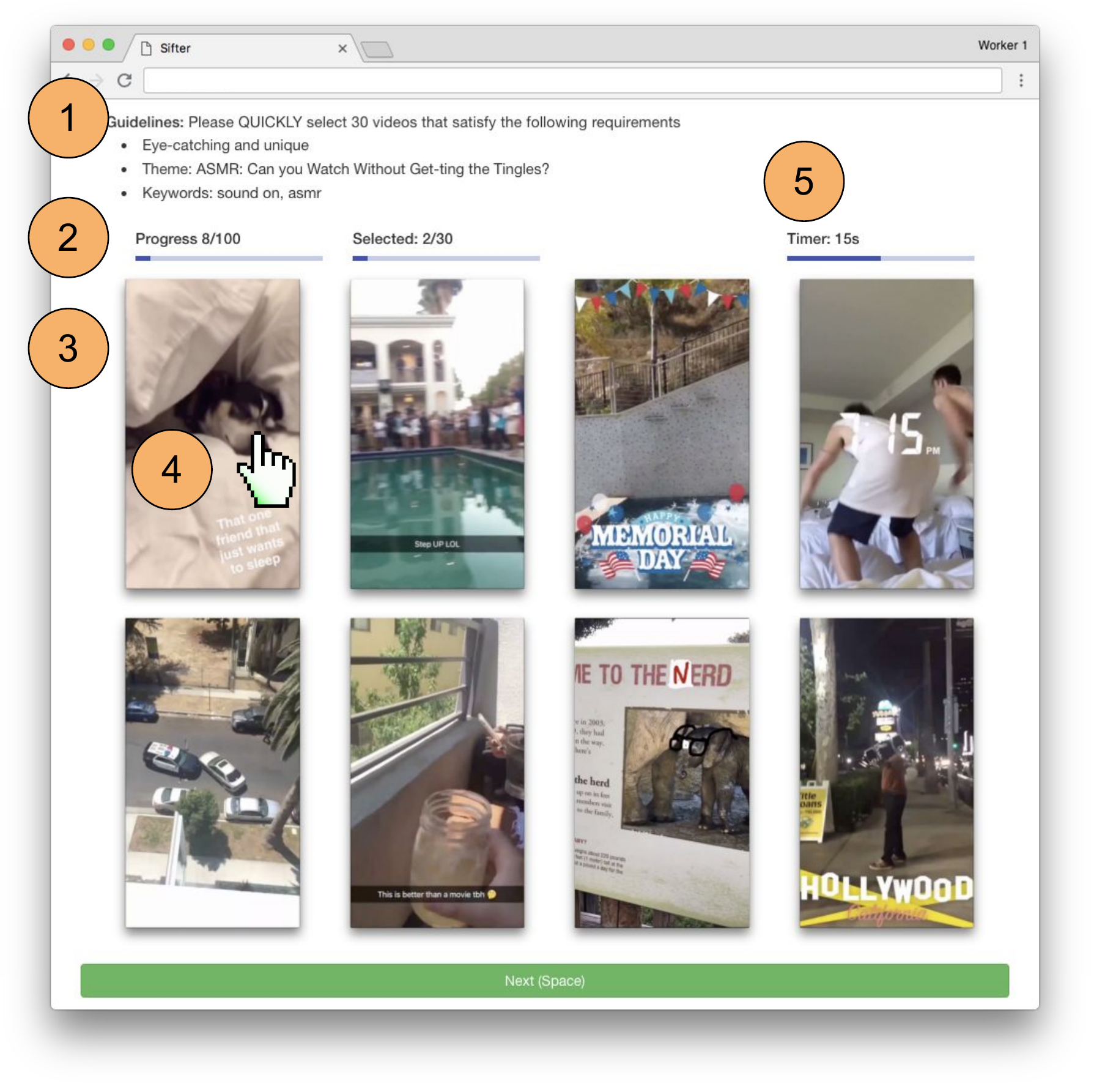}
  \caption{\sys{} UI task page. Workers can see the instructions, their progress (number of videos they have seen and selected), and a timer on every page. All videos autoplay silently when workers arrive on the page; they can mouse over videos to turn on audio, and click to select.}
  \label{final_ui}
\end{figure}

\subsubsection{Landing page with example-based instructions.}
Before a worker starts executing the task, they see a landing page where they get instructions (e.g., Please QUICKLY select \#\# videos that satisfy all the requirements.) and example videos from a previously published compilation (randomly selected). To make sure workers understood the task details, such as what HQ and TR videos are, we first tested their comprehension by providing different levels of contextual information in the instructions (see below a, b, c) and measured the quality of the outcome (rated by researchers) and the completion time.
We found that providing the goal of the task (a, b) and information about how the videos were found (c) increased the quality of selected videos without increasing the task time.



\subsubsection{Task page.}
The component index is corresponding to the numbers in Figure ~\ref{final_ui}. \textbf{1. Contextualized instructions.} 
These instructions reiterate what was presented on the landing page, but without the example videos. \textbf{2. Progress bar.} We used a progress bar to show workers how many videos are still needed, how many are left in the pool, and how many they have selected. \textbf{3. No scrolling.} To design a user interface that lets workers rapidly sift through videos, we first implemented a web interface with all videos on one page and asked workers to select interesting videos.
From follow-up interviews with workers, we found that displaying all the videos on a single page is inefficient because workers would forget what videos they had reviewed already as they scrolled up and down the page.
Thus we designed a layout to display as many videos as we could per page while avoiding having workers scroll. \textbf{4. Looping videos and audio on mouse over.} To enable fast visual scanning of the videos, each task page was populated with eight looping videos. These videos were muted, however, workers could move their mouse over any video to trigger its audio. This approach helped workers to rapidly go through a large corpus of videos.
Furthermore, we experimented with using keyboard shortcuts to play, pause, and select videos, but we found that workers were faster with the mouse-based approach. Also, using the mouse resulted in more videos being mouse hovered (reviewed) and selected.
We also experimented with different video preview speeds but found no difference in execution time compared to normal video speed. \textbf{5. Timer.} The sifting task is such that it is not necessary to select every good video, but only a small set of them. 
Additionally, as one of our goals is to speed up the process, we set a 30-second  limit on each page to prevent workers from getting stuck watching videos in great detail. 

\begin{table*}[t]
\centering
\resizebox{\textwidth}{!}{%
\begin{tabular}{p{1.8cm}|p{4.5cm}|p{3.5cm}|m{2cm}|m{2.5cm}|m{2cm}|m{2.5cm}}
\hline
{Compilation} & \textbf{Theme} &  \textbf{Keywords} & \textbf{\begin{tabular}[c]{@{}l@{}} Input videos \end{tabular}} & \textbf{\begin{tabular}[c]{@{}l@{}}Videos after \\ automated filters\end{tabular}} & \textbf{\begin{tabular}[c]{@{}l@{}}Workers \\ selection stage\end{tabular}} & \textbf{\begin{tabular}[c]{@{}l@{}}Workers \\ agreement stage \end{tabular}} \\ 
\hline\hline
{\textbf{C1}} &ASMR: Can You Watch Without Getting the Tingles? & sound on, asmr & 1,922 & 329 & 1 & 2 \\ \hline
{\textbf{C2}}&Black Panther: Does It Live Up to the Hype? & black panther & 1,984 & 586 & 1 & 2 \\ \hline
\textbf{C3}&Conspiracy Theories: 12 Videos That'll Make You Believe & ufo, ghost, conspiracy, alien & 3,262 & 1,262 & 2 & 2 \\ \hline
\textbf{C4}&Flashback Feels: Gone But Never Forgotten & flashback, throwback, 80s, 90s, 2000s & 2,118 & 980 & 1 & 2 \\ \hline
{\textbf{C5}}&Magic Wins: These Weird Tricks Will Fool You & magic, tricks & 3375 & 1,508 & 2 & 2 \\ \hline
{\textbf{C6}}&Fun Moms: They've Gone Wild... But We're Here For It & mom, mother, ma & 2,741 & 1,185 & 2 & 2 \\ \hline
\textbf{C7}&That's Nasty: 10 Videos That'll Make Your Skin Crawl & gross, disgusting, ew & 3,204 & 974 & 1 & 2 \\ \hline
\textbf{C8}&So Over It: Cue the Eye Roll \includegraphics[height=1em]{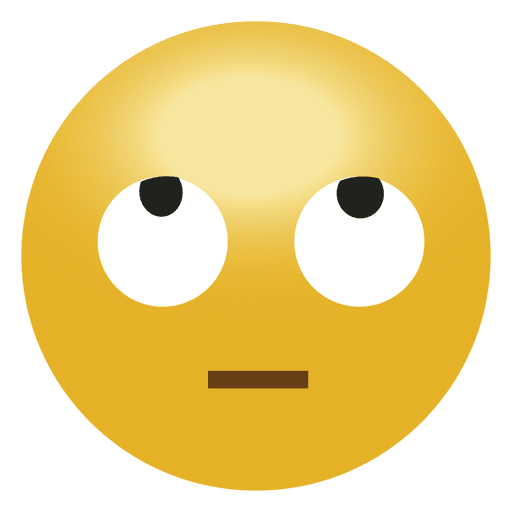} & so over it, i'm done, not amused, ugh, bummer & 3798 & 1,670 & 2 & 2 \\ \hline
\textbf{C9}&Happy St. Patrick's: Are You Ready to Shamrock \& Roll? & patricks's, patrick & 3,121 & 1,933 & 2 & 2 \\ \hline
\textbf{C10}&School's Out For Snow: The Weather's Got Us Wild & snow, school, campus, canceled & 847 & 426 & 1 & 2 \\ \hline
\textbf{C11}&Streaks: \includegraphics[height=1em]{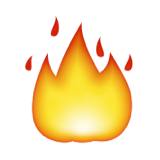}\includegraphics[height=1em]{figures/fire.png}\includegraphics[height=1em]{figures/fire.png}\includegraphics[height=1em]{figures/fire.png}\includegraphics[height=1em]{figures/fire.png} & streaks & 5,490 & 1,672 & 2 & 2 \\ \hline
\textbf{C12}&Weddings: These Brides \& Grooms Are the Real MVPs & wedding, weddings, brides, grooms & 2,206 & 1,390 & 2 & 2 \\ \hline
\end{tabular}
}
\caption{List of compilations used for evaluating \sys{}. All of these compilations had been previously published. Keywords were the words used by the platform's curators to find the videos to create the compilations. The number of videos retrieved using those keywords for the evaluation of \sys{} is in the fourth column. The fifth column presents the number of videos left after the automated filter step. The last two columns present the number of workers involved in each stage of filtering. 
}
\label{story_data}
\end{table*}

\section{Evaluation}

\sys{} was meant to scale subjective human judgment.
Also, we wanted \sys{} to perform fast and reliably. 
To this end, we compared the quality of the videos that \sys{} generated to those generated by dedicated curators. We report the details for each compilation we evaluated in Table~\ref{story_data}.

\subsection{Dataset}
To effectively validate the video quality, we selected 12 compilations that were previously created and published by staff curators by the partner platform. 
We put together these compilations with the following guidelines:
\begin{enumerate}
\item All videos were in English to make it easier for the researchers to run the evaluation. 
\item Each compilation had more than 1,000 videos to curate. 
\item The videos represented a wide range of themes from international events to evergreen topics. For example, ``ASMR,'' which contained videos with soothing sound, or ``That's Nasty'' with videos of showing a ruined ice cream.
\item The topics of the videos were globally recognizable, i.e., we avoided local news.
\end{enumerate}

For each compilation we gathered (a) the number of staff curators that worked on putting together those compilations, (b) the name or theme of the compilation, (c) the keywords used to find all the  videos that were considered for the compilations, and (d) the time spent searching for and collecting the videos.  Table~\ref{story_data} lists (b) and (c);  Table~\ref{timecompare} shows the aggregate of (a) and (d).

\subsection{Workers and Evaluators }

We recruited both the workers and evaluators from an online freelancing platform that enables workers to set their own rates. Although the platform allows requesters to bargain, we took the rate proposed by the workers at face value. 

\begin{itemize}

\item Workers. We recruited nine human workers (two female, seven male) whose self-reported expertise was ``data entry.'' Workers came from Europe, Asia, and North America. Any worker who applied to work on the task was accepted on a ``first come, first serve'' basis.
The last two columns in Table~\ref{story_data} report the number of workers involved per stage. 
\item Evaluators. We recruited three professional video producers (one female, two male) to evaluate the quality of the final videos. 
They came from North America and Asia.
These evaluators were chosen because they all had prior experience with the process for Snap video curation, but none of them were familiar with  the compilations we selected. 


\end{itemize}

\subsection{Procedure}

Once the workers received the task, they first read the aforementioned task instructions (e.g., name, example videos). 
Based on how many videos have been selected in the human worker pipeline, the workers were assigned to either the ``selection'' or the ``agreement'' stage in the pipeline.
After completing their tasks, workers filled out a survey about their familiarity with the topic of the compilation, the challenges they faced, and their selection strategy.

\begin{table*}[t]
  \centering
  \resizebox{0.8\textwidth}{!}{
  \begin{tabular}{c | c | c | c | c}
    \toprule
     & \begin{tabular}{@{}c@{}}Avg. workers (s.d.)\end{tabular}
     & \begin{tabular}{@{}c@{}}Avg. generated videos (s.d.) \end{tabular} 
     & \begin{tabular}{@{}c@{}}Avg. time spent (s.d.) per compilation\end{tabular} 
     & \begin{tabular}{@{}c@{}} Avg. time spent (s.d.) per video\end{tabular}
     \\
     \midrule
    \sys{}  & 3.58 (0.51) & 21.08 (8.36) & 13.55 min (5.50) & 0.71 min (0.41)  \\
    Curator  & 2.58 (1.56) & 120.67 (91.02) & 259.79 min (252.82) & 2.35 min (1.49) \\
    \bottomrule
    \end{tabular}
    }
    \caption{Comparing the production of workers using \sys{} and professional curators. The first column reports the average number of people involved per compilation. The second column presents the average number of videos generated per compilation. The third column reports the time spent per compilation. 
    }
    \label{timecompare}
\end{table*}

\section{Results and Discussion}

Overall, \sys{} is faster than the staff curators at generating a refined set of videos, and the quality of these videos is comparable to those identified by the curators. In this section, we discuss details of our study results.

\subsection{\sys{} is three times faster than curators.}
We computed the average time spent per compilation for \sys{}  by adding up the completion times of the two workers who took longer to finish R2 and R3 stages (Eq.\ref{avg_timespent}).
This is because the workers from the same stage can perform tasks in parallel.
The fourth column in Table~\ref{timecompare} reports the average time spent per compilation between two methods.
With that, we computed the average human-time spent per video selection between the two methods (last column in Table~\ref{timecompare}). We found that Sifter ($\mu = 0.71$min, $\sigma = 0.41$min) can pick a video three times faster than curators ($\mu= 2.35$min, $\sigma = 1.49$min) ($p < .0014$).

\begin{equation} \label{avg_timespent}
\begin{split}
    \text{\sys{}}_{T} & = \max(t_{\text{worker1,selection}}, t_{\text{worker2,selection}}) \\
     & +  \max(t_{\text{worker1,agreement}}, t_{\text{worker2,agreement}})
     \end{split}
\end{equation}

To determine the time that the curators spent sifting each compilation, we used the conventional timeout cutoff technique to determine their query session and then added up all the query sessions per compilation~\cite{jones2008beyond}. 
We used 30 minutes as the timeout threshold, meaning that if the next search request happened more than 30 minutes after the current one we counted it as a new query session. 
The sum of all the session time is the final time spent per compilation.
Additionally, because \sys{} works asynchronously, curators could perform multiple compilation sifting tasks simultaneously.


\subsection{No quality difference for 11 of the 12 compilations.}
We evaluated \sys{}'s quality by comparing its output against the output of the staff curators when performing the same ``selection and collection'' stage that \sys{} aimed to replace, (step 2 in the pipeline from Fig.~\ref{curation_pipeline}).  
We did not inform evaluators where the videos came from, so they did not know whether it came from \sys{} or the staff curators. 
We were interested in measuring how relevant the videos output by \sys{} were to the topic of the compilation. 
For example, for compilation C5, we wanted to know how relevant the videos selected by \sys{} were to the topic of ``Magic Wins: these weird tricks will fool you.''

\begin{figure}[b]
  \includegraphics[width=0.5\textwidth]{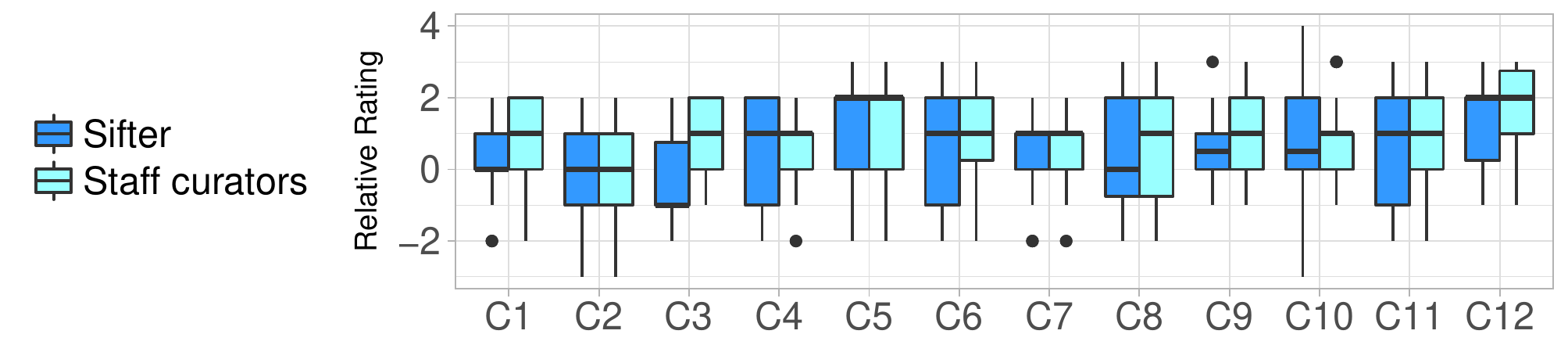}
  \caption{A box plot comparing \sys{} with staff curators using a relative rating measurement. For C3, the curator's ratings were significantly different from Sifter's ($p < .0042$). }
  \label{rating}
\end{figure}

For each compilation, we calculated a rating for a sample of videos output by \sys{}, and another for the ones output by the staff curators. We hired raters to evaluate each video in the sample on a 5-point Likert scale for ``how relevant is this video for the topic `insert topic'?'' from  1 (not relevant at all), to 5 (very relevant).

To reduce the biases resulting from the raters' different background knowledge (according to their feedback), we also added a baseline condition that consists of a sample of randomly selected videos.
These videos were retrieved from the corpus using the keywords described in Table 3, e.g., 10 random videos out of the 1,984 videos that were collected for compilation C2. 
Then we measured \sys{}'s and the staff curators' ratings relative to the baseline rating.

The rating for each video in the \sys{} sample was calculated by subtracting the average of the baseline ratings from the rating given by the rater to that video. We then calculated the average of all of the individual ratings in \sys{} and used that as the rating for \sys{} for that compilation, e.g., 0.93 for compilation C5 in Figure~\ref{rating}.
In this way, we took into account the individual raters' differences in perception and were able to make the difference comparison to determine the effectiveness of \sys{}.
We conducted 12 comparisons using two-tailed, paired-samples t-test, and with Bonferroni correction, we considered the comparison result significant if the $p-$value was below .05/12 = .0042. 
We found that the ratings for eleven out of the twelve compilations generated by \sys{} showed no significant differences ($p > .0042$); the other one compilation, C3, were rated significantly lower ($p < .0042$).
We analyzed the reasons in detail in a later section.





\subsection{Workers use different strategies.}

In order to improve the automation process in the future, we wanted to understand what strategies workers used when selecting videos. 
We then gave workers a questionnaire when they finished the selection task of each compilation.
The questionnaire contained free-text questions about their strategy such as ``what was your strategy for completing this task?''
We then coded their answers with one of a set of five categories that we came up with through an inductive approach inspired by the text instructions.

\begin{figure}[b]
 
  \includegraphics[height=3.5cm, width=0.5\textwidth]{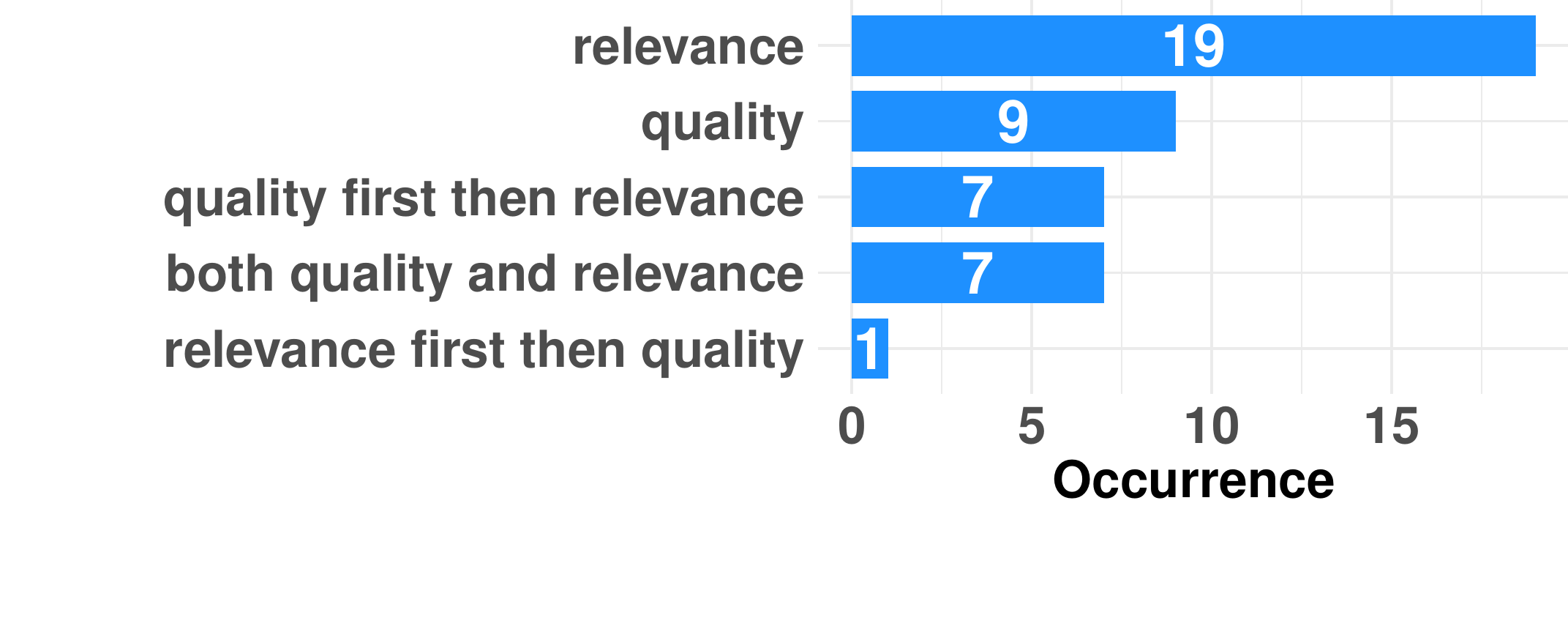}  
  \caption{Histogram of the popularity of different strategies among human workers.}
  \label{strategyoccur}
   \vspace{-1.5pc}
\end{figure}

Figure~\ref{strategyoccur} shows how often each strategy was used. 
We found that for more than half of the compilations (7/12) workers reported using at least two strategies, and two of them had workers report using only one (C2, C4).

A relevance-centric strategy means the worker reported focusing primarily on identifying videos that were relevant to the theme of the compilation. For instance, a worker mentioned ``I was looking for videos that match given theme and move on'' (P33).
A quality-centric strategy means that the worker reported having focused on identifying ``eye-catching'' or ``interesting'' videos. For example, a worker reported ``I just see which video is most interesting, look good, have a magic or some funny or eye catching things in it'' (P16).
We also found that some workers reported applying both strategies equally, e.g. ``Watching eye-catching videos and select them if they match given theme'' (P40), and others reported applying one strategy before the other, e.g. ``I was looking for interesting videos as fast as I could and then I was making selection if they are right match'' (P14).

\section{Limitations and Future Work}

\subsection{Parameter values in the pipeline}

One of our contributions is the design of \sys{}'s pipeline.
However, the parameter values we derived for the final evaluation were based on our need to form compilations with 10 to 20 videos. Future users of \sys{}'s pipeline would need to find their own optimal values parameter for their data and scenarios.
Future work should also explore the dynamics of the parameter values through additional controlled studies. 

\subsection{Evaluating Automated Filter}
Because we designed our automated filters (R1) using previously published and evaluated heuristics, we did not evaluate their performance as part of this paper. However, future work would benefit from a re-evaluation of these processes for each new context.

\subsection{Worker Biases}

Although we strove to understand what is ``universally'' relevant by ``averaging out'' workers' biases, \sys{} still serves as only a proxy and might not always perform as expected.
For example, there was one compilation, C3, in which no worker reported unfamiliarity with the theme, but \sys{} still generated videos that were given lower quality ratings than the curators' compilation.
We analyzed the outcome and found that the majority of videos in the set were selfies with an alien animation, which was briefly a popular camera effect.
Although the content is relevant to one of the keywords in the query `alien', it is not considered to be a high-quality video based on our definition, and also made the majority of the selection look similar in style. 
A worker in T8 commented that:

\begin{quote}
    \textit{[The] main challenge was to not pick all the same videos, cause [sic] there are many similar videos.}
\end{quote}

We further analyzed the comments given by workers for this compilation. One worker from the agreement stage reported that \textit{``There isn't that many eye-catching videos ...''}.
As we asked workers in the agreement stage to select at least 30 videos, they might have selected some with low quality because of this requirement.
One way to address this is to let workers self-report low quality video batches, especially in the agreement stage.
We could do this by building on prior research done on quality control \cite{dow2012shepherding}.

\subsection{From Curation to Moderation}
Our primary focus in this work was on video curation of interesting content; however, our approach is also promising for moderating inappropriate videos. For example, workers could potentially rapidly identify and remove videos of kids bullying or being engaged in violent activities. 
In the future, \sys{} has the potential to delight the users of online social media platforms with safe content.


\subsection{Opaque Strategies}
We have made initial attempts at trying to figure out what strategies curators and workers used for successful sifting. 
However, a more formal field study with curators to explore how their strategies change for different compilation themes could offer valuable insights.
For example, examining the portions of the video that people watched and the interaction patterns people have with videos might provide useful information.
In addition, we focused on the sifting task in this work as the first step. Future work can expand our approach to other steps, such as leveraging human workers for complex queries, to further scale the curation process.

\section{Ethical recommendations}
We believe the \sys{} approach has significant potential for being widely adopted, so it is important for us to ensure that designers who build upon this work use it ethically. 
During the curation process, we expect curators to have their own control in the workflow--- 
e.g., taking breaks when needed, driving their own work forward, and selecting videos that are potentially interesting to them. 
To ensure they are treated ethically and responsively, we include the following recommendations on how to  appropriately deploy \sys{}:

\begin{itemize}
    \item \textbf{Compensation.} Workers should be paid a fair rate~\cite{fairwork}. 
    \item \textbf{Sessions.} Session lengths should be capped, and breaks should be compensated to care for workers' mental and physical health. 
    \item \textbf{Choice.} Workers' sensibilities and personal preferences should determine which topics they curate. We envision a scenario where workers get to see a list of titles and descriptions of stories they can curate, giving them the power to select which ones they work on. It is also important to understand their familiarity with the topics, as it would make them comfortable pursuing a task and improve the quality of work. 
    \item \textbf{Transparency.} End-users who consume or produce content curated by a hybrid process like \sys{} should be informed of the process by which the stories are curated. 
\end{itemize}

\section{Conclusion}
In this paper, we introduced \sys{}, a system that utilizes automation and workers to enable curators to delegate the task of selecting eye-catching and thematically-relevant videos, allowing them to focus on more creative tasks. 
\sys{} first leverages video processing techniques to remove unqualified videos, and then uses a human-powered pipeline that allows workers to rapidly browse, select, and reach an agreement on videos. 

We evaluated \sys{} by creating 12 different video compilations, and found that the quality of the majority of those compilations was indistinguishable from the ones created by staff curators. 
We believe that our findings can inform the design of systems in the future that rely on subjective human judgments at scale.

\section{Acknowledgements}
Special thanks to our colleagues Maarten Bos, Kelly Mack, and Aletta Hiemstra for their feedback. 

\bibliographystyle{ACM-Reference-Format}

\balance{}
\bibliography{sample}

\end{document}